%% file: manuscript.tex
\documentclass[twocolumn]{aastex63}

\usepackage{xspace}
\usepackage{amsmath}
\graphicspath{{figures/}}

\newcommand{\Kepler}{\textit{Kepler}\xspace}

\newcommand{\abs}[1]{\lvert #1 \rvert}
\newcommand{\ZZ}{\ensuremath{\mathcal{Z}}\xspace}
\newcommand{\Zcom}{\ensuremath{\mathcal{Z}_\mathrm{com}}\xspace}
\newcommand{\Zcross}{\ensuremath{\mathcal{Z}_\mathrm{cross}}\xspace}
\newcommand{\Zunstable}{\ensuremath{\mathcal{Z}_\mathrm{unstable}}\xspace}

\begin{document}
%% Front Matter
\title{How Close are Compact Multi-Planet Systems to the Stability Limit?}

\author[0000-0001-7961-3907]{Samuel W.\ Yee}
\affiliation{Department of Astrophysical Sciences, Princeton University, 4 Ivy Lane, Princeton, NJ 08540, USA}
\author[0000-0002-9908-8705]{Daniel Tamayo}
\affiliation{Department of Astrophysical Sciences, Princeton University, 4 Ivy Lane, Princeton, NJ 08540, USA}
\author[0000-0002-1032-0783]{Samuel Hadden}
\affiliation{Harvard-Smithsonian Center for Astrophysics, 60 Garden St., MS 51, Cambridge, MA 02138, USA}
\author[0000-0002-4265-047X]{Joshua N.\ Winn}
\affiliation{Department of Astrophysical Sciences, Princeton University, 4 Ivy Lane, Princeton, NJ 08540, USA}

\begin{abstract}
Transit surveys have revealed a significant population of compact multi-planet systems, containing several sub-Neptune-mass planets on close-in, tightly-packed orbits.
These systems are thought to have formed through a final phase of giant impacts, which would tend to leave systems close to the edge of stability. 
Here, we assess this hypothesis, comparing observed eccentricities \replaced{as measured by}{in systems exhibiting} transit-timing variations (TTVs), with the maximum eccentricities compatible with long-term stability.
We use the machine-learning classifier SPOCK \citep{Tamayo2020} to rapidly classify the stability of numerous initial configurations and hence determine these stability limits.
While previous studies have argued that multi-planet systems are often maximally packed, in the sense that they could not host any additional planets, we find that the existing planets in these systems have measured eccentricities below the limits allowed by stability by a factor of 2--10.
We compare these results against predictions from the giant impact theory of planet formation, derived from both $N$-body integrations and theoretical expectations that in the absence of dissipation, the orbits of such planets should be distributed uniformly throughout the phase space volume allowed by stability.
We find that the observed systems have systematically lower eccentricities than this scenario predicts, with a median eccentricity about 4 times lower than predicted.
These findings suggest that if such systems formed through giant impacts, then \replaced{some orbital damping must have occurred during or after formation, perhaps due to interactions with the natal gas disk or a leftover population of planetesimals.}{some dissipation must occur to damp their eccentricities.
This may take place during formation, perhaps through interactions with the natal gas disk or a leftover population of planetesimals, or over longer timescales through the coupling of tidal and secular processes.}
\end{abstract}

\section{Introduction}
One of the major findings of the \Kepler space telescope was the discovery of numerous exoplanet systems containing multiple planets with masses between the masses of Earth and Neptune, on orbits that are close-in and tightly-packed.
The final assembly of these planets is thought to begin with smaller rocky bodies of mass $\sim 0.1 - 1.0\,M_\oplus$, which either form {\it in situ} close to the host star, or migrate inward from a formation location further out in the protoplanetary disk (see e.g.~review by \citealt{Raymond2014}).
These protoplanets are initially on nearly circular and coplanar orbits, but as the gas disk dissipates, their orbits get mutually excited, leading to a ``giant-impact'' phase where bodies collide and grow to their final planetary masses.

In this picture, the orbital elements of protoplanets diffuse chaotically due to mutual gravitational interactions, eventually arriving at an unstable state, when a collision or scattering event occurs.
Following this instability, the system is left in a new quasi-stable configuration, only for this process to repeat itself.
At each stage, the planetary system would have a stability timescale comparable to its age (e.g. \citealt{Laskar1996}).
The result of such an evolutionary process would be a population of planetary systems with a distribution of orbital properties that extends all the way up to their respective dynamical stability limits.
This could explain why many observed \Kepler multi-planet systems appear to be dynamically packed, with no space between planets to squeeze in an additional body (e.g. \citealt{Barnes2004a,Barnes2004,Fang2013}).

\cite{Pu2015} investigated this hypothesis that orbital architectures of systems are sculpted by dynamical instability, using $N$-body simulations of compact multi-planet systems.
They found that stability on Gyr timescales requires effective mutual Hill separations of $K \gtrsim 10$ for planets on circular orbits, increasing to $K \gtrsim 12$ if a small amount of eccentricity and non-coplanarity are introduced.
When they compared this to the observed Hill separation distributions of \Kepler systems, they found a mean effective spacing of $K_\mathrm{obs} = 12.3$, which they took to be evidence that the population of such systems does indeed extend to the edge of stability.

An important difficulty with such studies is that instability times are strong functions of interplanetary separation \citep{Chambers1996}, orbital eccentricity \citep{Zhou2007} and proximity to nearby resonances \citep{Obertas2017}.
This implies that previous approaches of drawing orbital elements independently to generate synthetic multiplanet system populations can miss strong correlations between these parameters.
A better approach would be to model individual observed systems to capture these dependencies.

The first main challenge to such an approach is observational. While most discovered compact systems transit their host star and thus have precisely measured orbital periods, it is difficult to constrain their orbital eccentricities and pericenter orientations.
Radial-velocity measurements have either not been undertaken, or are not precise enough to constrain the eccentricity.
Transit duration measurements are typically only able to place weak upper limits on planet eccentricities, on the order of $\sim 0.1$ (e.g. \citealt{VanEylen2015}).
Additionally, the masses of these planets, which also affect their stability, are often poorly-determined from radial-velocity measurements due to these planets' typically small sizes and the faintness of their host stars.

Fortunately, there is a subset of compact planetary systems for which we have much more precise knowledge of both masses and eccentricities: systems exhibiting transit-timing variations (TTVs).
In such systems, the strong perturbations between planets close to mean motion resonances (MMRs) lead to measurable changes in planetary transit times, from which we can infer their masses and orbital parameters.
\cite{Hadden2017} used the TTV catalog of \cite{Rowe2015} to derive masses and eccentricities of 145 planets, with a typical mass precision of a few times 10\%.
They also found that most of these planets have small but nonzero eccentricities (median $e \sim 0.02$), possibly the remnant eccentricities that were produced during the giant-impact phase.
The planets in these TTV systems are amongst the only planets of sub-Neptune size and smaller with individual eccentricity and mass constraints, providing us with a good sample for studying the limits dynamical stability places on eccentricities.

The second roadblock to such an analysis is computational. $N$-body integrations for even $10^9$ orbits, which typically represents $\sim 0.1-10\%$ of these systems lifetimes, requires several hours of CPU time with current hardware. 
This renders a multi-dimensional exploration of an observed multi-planet system's phase space extremely computationally challenging. 
To alleviate this problem, we use the Stability of Planetary Orbital Configurations Klassifier (SPOCK), a machine-learning model developed by \cite{Tamayo2020} to classify the long-term stability of compact planetary systems.
SPOCK works by running a short trial integration of the system for $10^4$ orbits of the innermost planet, monitoring a number of dynamical variables over this short run.
It then uses these quantities to predict the probability that the system will be stable for $10^9$ orbits.
Since this final step is fast compared to the trial integration, SPOCK provides a $\sim 10^5$ speedup factor in the time required to evaluate the stability of a given set of initial conditions.
\cite{Tamayo2020} trained SPOCK on near-resonant systems like the ones in our sample, and found that it generalized well to all compact multi-planet systems, outperforming simpler stability criteria such as Hill stability \citep{Gladman1993} and AMD stability \citep{Laskar2017}.
Using this new tool, we are able to explore the multi-dimensional eccentricity phase space more rapidly and comprehensively than if we were to use a traditional $N$-body integration.

In this paper, we focus our investigation on TTV systems containing 3 or more planets with good mass and eccentricity constraints, which we describe in Section \ref{sec:systems}.
Our experiment involves increasing the measured eccentricities of the planets in this sample until they go unstable, allowing us to measure how far the observed systems are from the stability limit (Section \ref{sec:alpha_ecc}).
We then try to interpret these results in the context of the giant-impact planet formation theory using $N$-body simulations and theoretical work to predict how the parameters of planetary systems should be distributed (Section \ref{sec:giant_impact}).
We discuss some caveats in Section \ref{sec:discussion}, and conclude in Section \ref{sec:conclusion}.

\section{TTV Systems \label{sec:systems}}

We draw the systems for our study from the analysis of \cite{Hadden2017}, hereafter H17, who derived masses and eccentricities for 55 planetary systems based on the \Kepler TTV catalog of \cite{Rowe2015}.
For a system of two planets, there is an analytic stability criterion that marks a relatively sharp transition between long-lived configurations and those that rapidly go unstable (e.g. \citealt{Deck2013,Hadden2018}).
Such systems are less likely to undergo the constant rearrangements which would sculpt their architectures.
We therefore use only systems containing three or more planets whose properties could be constrained by TTVs, which reduces the list to 22 systems.

Of the systems containing three or more planets, Kepler-60, Kepler-80, and Kepler-223 appear to be in resonant chains, as indicated by H17.
Mapping out the stable phase space of such systems is difficult because their long-term stability requires specific combinations of eccentricity and the apsidal angles, the latter of which are poorly constrained \citep[e.g.][]{Tamayo2017}.
Most ($\gtrsim 99\%$) of the initial configurations drawn from the measured posterior distributions are classified as unstable by SPOCK, an observation also made by H17 using $N$-body integrations.
Therefore, these resonant-chain systems appear to be substantially different from the other TTV systems, which are merely close to resonance but not truly in resonance.
Furthermore, the planets in resonant chains may have had different formation histories compared with the others, given that migration must have played an important role in moving them to their stable resonant configurations \citep[e.g.][]{Morrison2020}.
These considerations lead us to exclude these resonant systems from our analysis and focus on the remaining 19 systems.

Some of these systems have additional planets for which the properties are not constrained by the TTV analysis, because they do not exert a sufficiently strong influence on the neighboring planets.
In all cases, these non-TTV planets are either interior or exterior to all of the TTV planets in that system, with period ratios of $P'/P > 2.2$ relative to the closest planets.
Given that the TTV planets in the same systems are more tightly packed and have smaller period ratios, the overall stability of the system is unlikely to be strongly affected by the presence of these additional non-TTV planets. 
In our primary analysis, we ignore these additional planets, although we do check on the validity of this assumption in Section \ref{ssec:additional_planets}.
In total, we are left with 68 planets in 19 systems. 
These planets all have mass and eccentricity measurements from TTV analysis, with typical uncertainties of $\sim 30-50\%$.
We list the planets used in our analysis and their properties in Table \ref{tab:system_data}.

\begin{deluxetable*}{lcccccc}
\tablecolumns{7}
% \tablewidth{\columnwidth}
\tablecaption{TTV Planet Properties \label{tab:system_data}}
\tablehead{
	\colhead{Planet} & \colhead{Period} & \colhead{$M_\star$} & \colhead{$M_p$} & \colhead{$\abs{\ZZ}$} & \colhead{$\abs{\ZZ_\mathrm{com}}$} & \colhead{$\abs{\ZZ/\Zunstable}$} \\
	 & \colhead{(days)} & \colhead{($M_\odot$)} & \colhead{($M_\oplus$)} &  &  & 
}
\startdata
\input{tables/table1_stub.tex}
\enddata
\tablecomments{List of TTV systems and planets used in our analysis. Stellar and planetary masses, as well as free eccentricity \ZZ, were derived by \cite{Hadden2017}.
Because the free eccentricity \ZZ is a property of adjacent pairs of planets, we have recorded \ZZ in the row of the inner planet of the pair.
\added{The ``center-of-mass'' eccentricity for each system, $\abs{\Zcom}$ is computed according to Eq. \ref{eq:Zcom}, and recorded in the row of the innermost planet.}
$\abs{\ZZ/\Zunstable}$ is computed numerically in Section \ref{sec:alpha_ecc} and reflects the fractional distance to instability of the entire system.
The values and uncertainties reflect the mode of the posterior probabilities and 68.3\% highest posterior density intervals around the mode, or 68.3\% upper limits if this interval is consistent with zero.
This table is published in its entirety in the machine-readable format. A portion is shown here for guidance regarding its form and content.}
\end{deluxetable*}

\section{Eccentricity Limits for Multi-Planet Systems \label{sec:alpha_ecc}}

A straightforward test of whether multi-planet systems have eccentricities close to the stability limit is to ask: by what factor could we increase the eccentricities of observed systems before they become unstable on astrophysically short timescales?
The planetary systems in our sample all have small but non-zero eccentricities.
If these systems are on the edge of stability, then we might expect that a small increase in eccentricity would be sufficient to render them unstable.
Conversely, if we are able to increase their eccentricities by a large factor (e.g. $10$x) without endangering their stability, we would conclude that these systems are over-stable.

However, this experiment cannot quite be performed in the simple manner just described, because TTV observations do not lead to tight constraints on all of the relevant eccentricity parameters.  Rather, there are particular combinations of the eccentricities that are tightly constrained, and others that are poorly constrained.
It is therefore valuable to frame our numerical experiment in terms of the resonant variables that drive the TTV dynamics.

\subsection{Resonant Variables} \label{sec:resvar}
For a single pair of planets, we define the inner planet's complex eccentricity vector $z = e\exp{i\varpi}$ in the complex plane, where $e$ is the orbital eccentricity and $\varpi$ the longitude of pericenter, with a similar definition $z'$ for the outer planet.
Near a first-order MMR (with period ratios near $n:n-1$), one can show that to leading order, only a particular combination of eccentricities varies \citep[e.g.,][]{Sessin1984, Deck2013, Batygin2013},
\begin{equation}
\mathcal{Z} = \frac{f z + g z'}{\sqrt{f^2 + g^2}} \label{eq:Z},
\end{equation}
where $f$ and $g$ are order-unity coefficients determined by the particular MMR.\footnote{While Eq.\:\ref{eq:Z} strictly only applies at leading order near first-order MMRs, \cite{Hadden2019} shows that it also approximately holds true both at higher order in the equations of motion near first-order MMRs as well as for higher order MMRs for suitably chosen values of $f$ and $g$.} 
Apart from the 2:1 MMR, the coefficient $f \approx -g$, so $\mathcal{Z} \approx (z' - z)/\sqrt{2}$ \citep{Deck2013,Batygin2013}.
The combination $\mathcal{Z}$ thus approximately represents the {\it anti-aligned} component of the eccentricity.
In particular, aligned orbits can have large, nearly equal eccentricities and still have $\mathcal{Z} \approx 0$.

By contrast, there is a complementary combination that is approximately conserved \citep{Sessin1984, Batygin2013, Deck2013}, and which approximately represents a ``center-of-mass" eccentricity \citep{Tamayo2020}
\begin{equation} \label{eq:Zcompair}
\mathcal{Z}_\mathrm{com} \approx \frac{\mu z+ \mu' z'}{\mu + \mu'},
\end{equation}
where $\mu$ and $\mu'$ are the mass ratios of the inner and outer planets relative to the star.

To leading order, it is only $\mathcal{Z}$ and not $\mathcal{Z}_\mathrm{com}$ that affects TTVs.
Therefore, only $\mathcal{Z}$ can be precisely measured by analyzing TTV observations \citep[e.g.,][]{Lithwick2012}.
By contrast, TTVs do not strongly constrain $\mathcal{Z}_\mathrm{com}$.
Additionally, and again to leading order, the same combination $\mathcal{Z}$ determines the stability of compact two-planet configurations \citep{Deck2013, Hadden2018}.
While analytic stability predictions for eccentric, compact 3+ planet systems are not yet known, \cite{Tamayo2020} found that the oscillation amplitudes of $\mathcal{Z}$ for each adjacent pair of planets were important predictors of long-term system stability in their machine-learning model SPOCK.
The fact that TTV observations specifically constrain the combination of eccentricities that strongly affects the stability of the system makes an analysis in terms of $\mathcal{Z}$ more illuminating than one in terms of the individual eccentricities.

We note that the evolution of $\mathcal{Z}$ can be approximately decomposed into a forced component determined by the system's proximity to resonance, and a free component that oscillates about the forced equilibrium point \citep[e.g.,][]{Deck2013}.
For the systems we examine, the forced eccentricity is typically at least an order of magnitude smaller than the free eccentricity, and $\mathcal{Z} \approx \mathcal{Z}_\mathrm{free}$.
For simplicity, then, we will refer to the values of $\mathcal{Z}$ as free eccentricities for the remainder of the paper (see e.g. \citealt{Lithwick2012,Hadden2017,Tamayo2021}).

Generalizing from a single pair to an $N$-planet system, there are $N-1$ free eccentricities, one for each pair of adjacent planets. 
Since we are only analyzing the near-resonant TTV systems, it is a good approximation to analyze the TTVs as a linear superposition of the interactions of adjacent planet pairs, rather than considering the effects of each planet on all other planets \citep[e.g.,][]{Hadden2017}.
This leaves only one remaining degree of freedom.
We choose to define a total center-of-mass eccentricity \citep{Tamayo2021},
\begin{equation}\label{eq:Zcom}
\mathcal{Z}_\mathrm{com} \equiv \frac{\sum_{i=1}^N \mu_i z_i}{\sum_{i=1}^N\mu_i}.
\end{equation}
This quantity is not necessarily conserved in a general near-resonant 3+ planet system, but in the limiting case where a single adjacent pair of planets dominates the mass, this quantity approaches the approximate analytic result given in Eq.\:\ref{eq:Zcompair}.

We focus on this set of dynamical variables $\{\ZZ_i, \ZZ_\mathrm{com}\}$ in the rest of our experiments, which we describe in the next subsection.

\subsection{Experimental Setup}

\begin{figure}
\epsscale{1.2}
\plotone{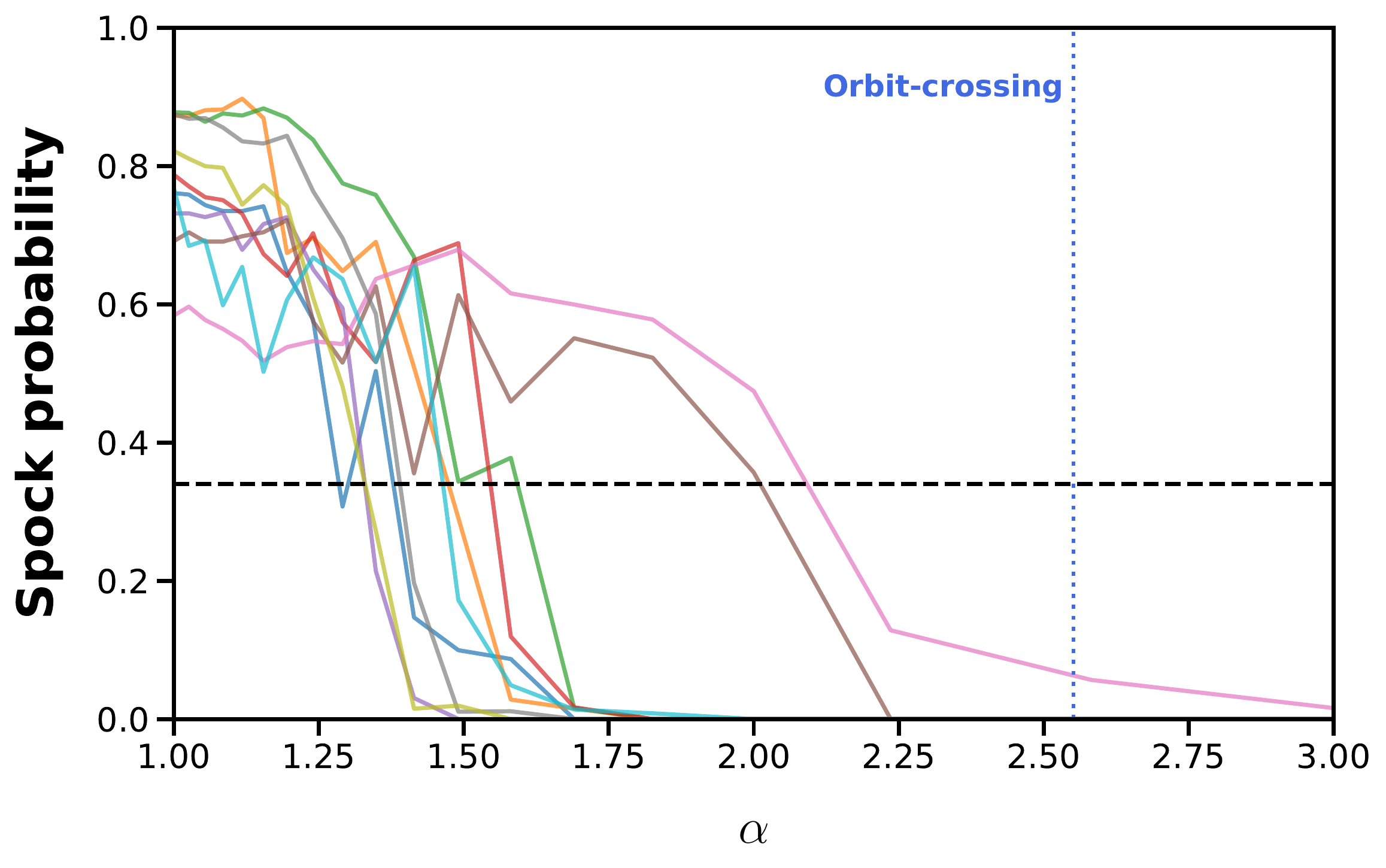}
\caption{Variation of the SPOCK probability of stability as the eccentricities of the planets are increased by a factor $\alpha$.
This plot shows an example for ten posterior rows, for the Kepler-52 system.
The horizontal dashed line shows the threshold probability of 0.34, but we found that varying this level did not significantly affect our results due to the relatively sharp transition between stable and unstable configurations.
In cases for which the probability of stability crossed the threshold more than once, we use the lowest value associated with the crossing of the threshold.
\label{fig:linear_alpha_example}}
\end{figure}

In this investigation, we increase all of the measured free eccentricities in the system, $\ZZ_{i,\mathrm{measured}}$ by the same factor $\alpha$, and find the threshold value that leads to instability.
Typically, there is a particular pair of planets whose free eccentricity renders the system most susceptible to instability.
Finding the threshold $\alpha$ beyond which the system is unstable is therefore akin to measuring $\abs{\Zunstable/\ZZ_\mathrm{measured}}$ for this most vulnerable pair of planets.

For each system, we draw 600 samples from the posterior distributions from the Markov Chain Monte Carlo TTV analysis of H17.
This marginalizes over the observational uncertainties in the masses and eccentricities.
We then multiply each $\ZZ_i$ by a factor $\alpha$, where $\alpha$ is sampled uniformly in log space from 1 to 20. 

TTV measurements often constrain $\abs{\ZZ_i}$ to within $10^{-2}$, whereas the uncertainty in $\abs{\Zcom}$ is on the order of 0.05 and can range up to 0.2.
While \Zcom is not expected to strongly influence stability \citep[e.g.,][]{Hadden2018}, such large uncertainties on \Zcom  mean that even $\alpha$ factors of 2--3 can lead to highly eccentric orbits, at which point strong nonlinear effects can quickly drive instabilities.
Therefore, we choose to hold \Zcom fixed at the corresponding value measured by H17, and only increase $\mathcal{Z}_i$.
This step distinguishes our procedure from simply scaling all the individual planet eccentricities $z_i \rightarrow \alpha z_i$, which would also increase \Zcom.

Our $N$ transformations
\begin{align}
\mathcal{Z}_i &\rightarrow \alpha\mathcal{Z}_i \qquad i=1,\cdots,N-1
\label{eq:Z_transform} \\
\mathcal{Z}_\mathrm{com} &\rightarrow \ZZ_\mathrm{com}.
\label{eq:Zcom_transform}
\end{align}
can be used to solve for the individual planet eccentricity vectors $z_i$.
Given that all of our systems are multi-transiting systems, the mutual inclinations are probably quite small; we assume them to be coplanar for simplicity (see Sec.\:\ref{sec:inc} for further discussion on this approximation).
Finally, we choose mean anomalies $M$ for each planet uniformly in $[0, 2\pi)$.
This procedure yields approximately $6\times 10^5$ initial planetary configurations (19 \Kepler systems $\times$ 600 posterior rows $\times$ 50 values of $\alpha$).

Then, we use SPOCK \citep{Tamayo2020} to classify the stability of each of these configurations.
For each posterior sample, we monitor the probability returned by SPOCK as we increase the eccentricity scaling factor, $\alpha$, as shown in Figure \ref{fig:linear_alpha_example}.
We record the value of $\alpha = \alpha_\mathrm{unstable}$ for which the SPOCK probability $\mathrm{Pr} < 0.34$, a threshold corresponding to a false positive rate of 10\% based on the training dataset of $N$-body integrations by \cite{Tamayo2020}. 
We experimented with different choices for this probability threshold and found that our results did not depend
sensitively on the choice. For a system defined by a given posterior sample from the TTV analysis, the transition between stable and unstable configurations is a relatively sharp function of $\alpha$ (Figure \ref{fig:linear_alpha_example}).
In cases for which the probability of stability crosses our threshold more than once as $\alpha$ increases, we choose the lowest value associated with the crossing of that threshold (e.g., blue curve in Fig.\:\ref{fig:linear_alpha_example}).
This yields a distribution of threshold $\alpha$ values for each system.
The quantity $1/\alpha_\mathrm{unstable} = \abs{\ZZ/\Zunstable}$ is then the fractional distance to stability; for example, $\abs{\ZZ/\Zunstable }= 0.5$ means the system could have free eccentricities larger than the measured eccentricities by a factor of 2, without being classified as unstable by SPOCK.

\subsection{Results} \label{ssec:alpha_results}

\begin{figure*}
\epsscale{1.1}
\plottwo{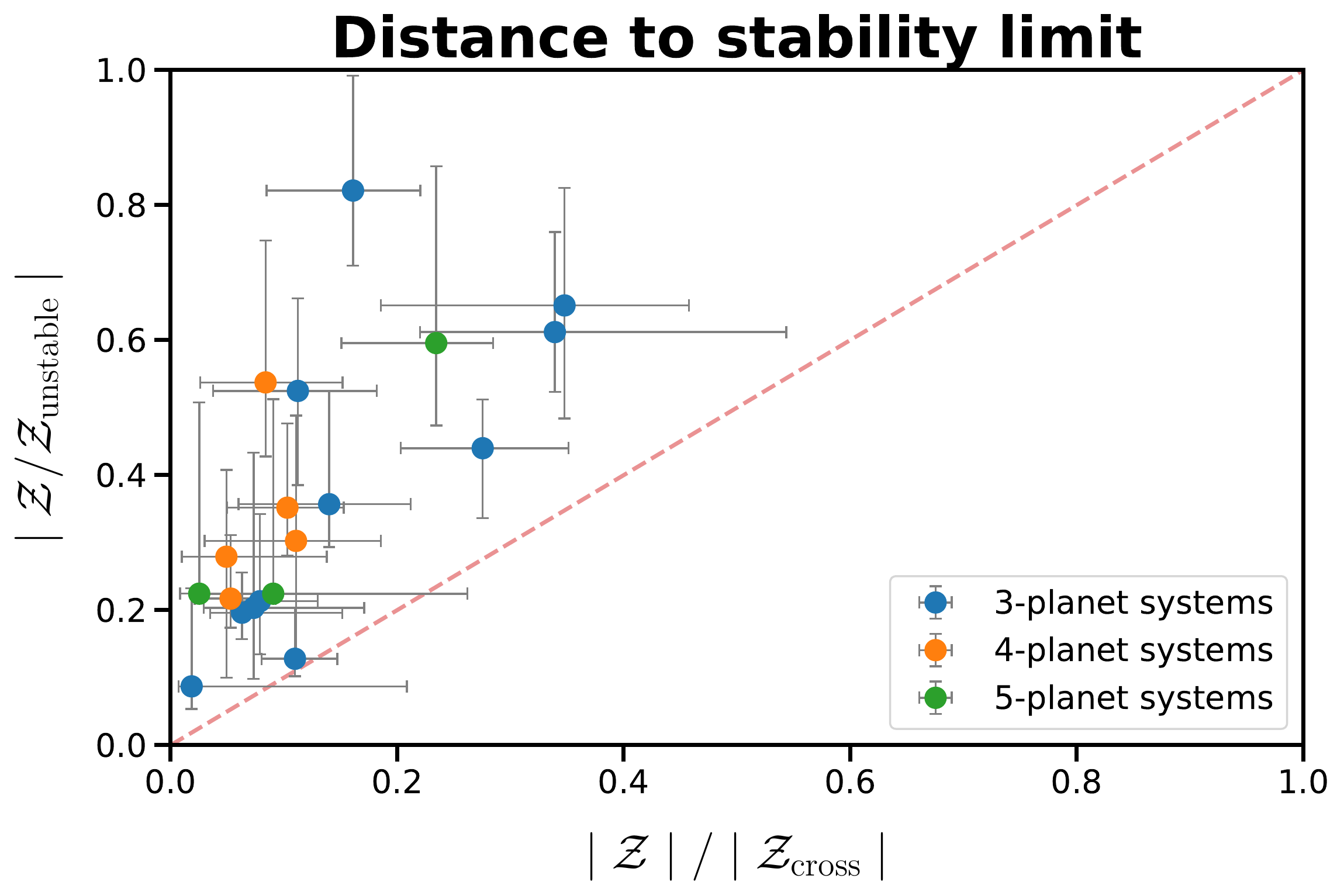}{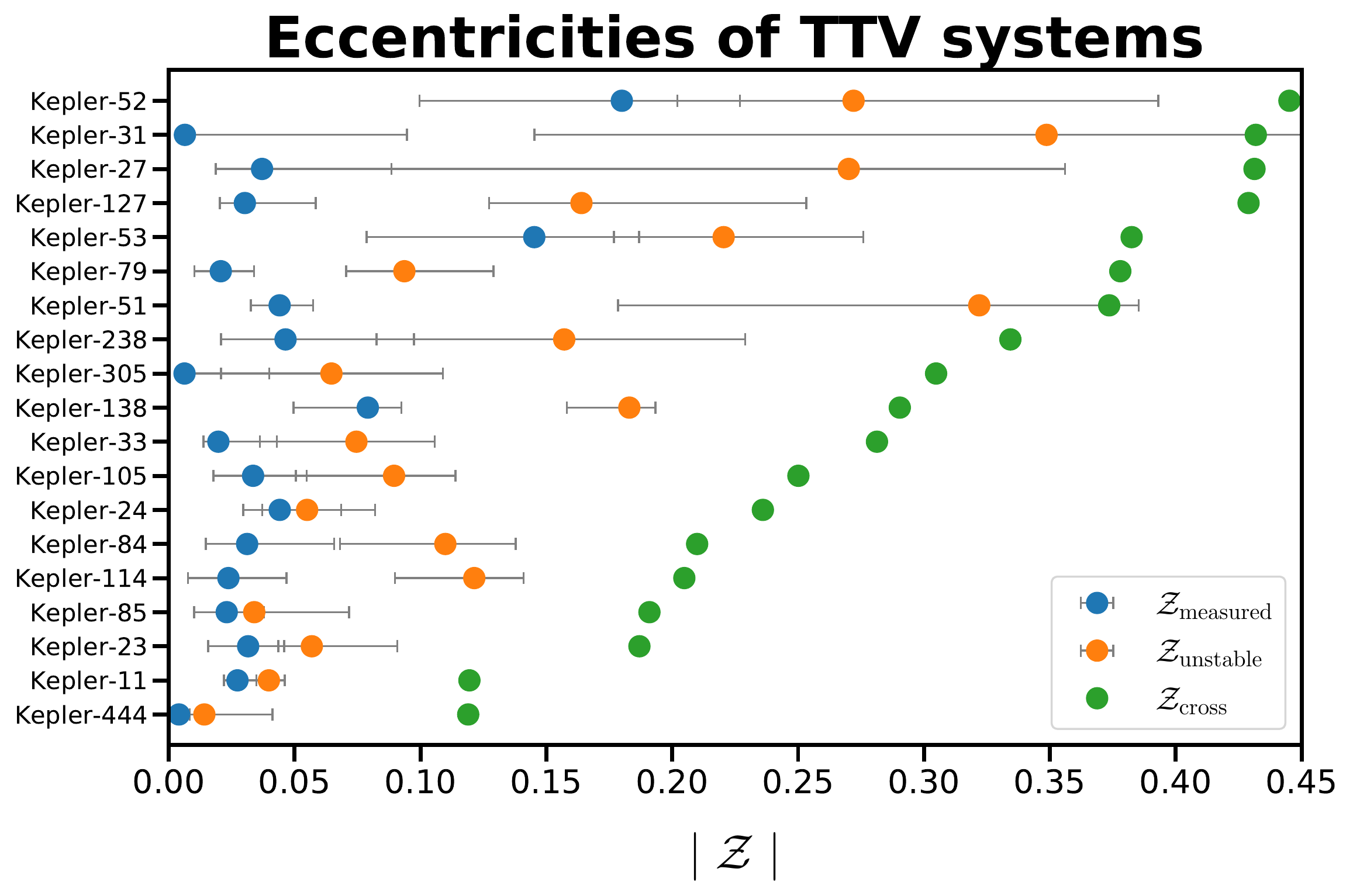}
\caption{\textbf{Left}: The fractional distance of a system to the stability limit, $1/\alpha_\mathrm{unstable} = \abs{\ZZ/\Zunstable}$ for each of the \Kepler TTV systems.
The x-axis shows the largest measured $\abs{\ZZ}/\abs{\Zcross}$ in the system.
All systems lie above the one-to-one line (red dashed line), reflecting the fact that the dynamical stability limit is more stringent than the requirement to avoid orbit-crossing.
Error bars in this and subsequent plots show the 68\% highest-density posterior intervals, while central points denote the mode of the respective distributions.
\textbf{Right}: The largest measured \added{eccentricity magnitude} $\abs{\ZZ}$ for a pair of planets in the system (blue), \replaced{along}{compared} with the corresponding $\abs{\ZZ_\mathrm{unstable}}$ at which the system goes unstable, \replaced{as well as}{and the orbit-crossing eccentricity} $\abs{\Zcross}$ (Eq. \ref{eq:Zcross}).
\label{fig:distance_to_stability}
}
\end{figure*}

The results of $\abs{\ZZ/\Zunstable}$ for each system are recorded in Table \ref{tab:system_data}, and plot in the left panel of Figure \ref{fig:distance_to_stability}.
The vertical axis shows $\abs{\ZZ/\Zunstable}$, while the horizontal axis indicates how far the system is from an orbit-crossing configuration, i.e., the largest value of $\abs{\ZZ}/\abs{\ZZ_\mathrm{cross}}$ among all planet pairs, where
\begin{equation} \label{eq:Zcross}
\abs{\ZZ_\mathrm{cross}} = \frac{1}{\sqrt{2}}\frac{a'-a}{a}
\end{equation}
is the orbit-crossing eccentricity to leading order in $\mathcal{Z}$ \citep[e.g.,][]{Hadden2019}.

As expected, none of the systems lie below the 1-1 line in Figure \ref{fig:distance_to_stability}; as eccentricity increases, they become unstable before the eccentricities are large enough to cause orbits to cross.
Systems with large measured eccentricities (large $\abs{\ZZ}/\abs{\ZZ_\mathrm{cross}}$) tend to be closer to the stability limit ($\abs{\ZZ/\Zunstable} \gtrsim 0.5)$.
However, at lower values of $\abs{\ZZ}/\abs{\ZZ_\mathrm{cross}}$, the systems appear to be overstable by factors that span a wide range.  Some systems, such as Kepler-24, are found within $20\%$ of the stability limit.  Others have eccentricities that are 10 times smaller than necessary to preserve dynamical stability.

If we assume that the planet pair with the largest measured eccentricity $\ZZ$ is responsible for driving the instability in each case, we can then estimate $\Zunstable$, the largest value of the free eccentricity compatible with stability.
In the right panel of Figure \ref{fig:distance_to_stability}, we plot \added{the magnitude of} this quantity for each system, where the systems have been ordered by $\abs{\Zcross}$. 
Again, we see that the spacing between the measured $\abs{\ZZ}$ in a system (blue points) and the $\abs{\Zunstable}$ (orange points) varies significantly between systems.
Another notable feature is that for systems with large $\abs{\Zcross}$, the stability limit $\abs{\Zunstable}$ approaches $\abs{\Zcross}$, but as the systems become more tightly packed, the stability limit shrinks more quickly than $\abs{\Zcross}$.
\cite{Hadden2018} found a similar result in the two-planet case: the onset of chaos due to resonance overlap occurs at lower values of $e/e_\mathrm{cross}$ as systems become more tightly spaced.
While those prior results are not directly applicable to our situation given the higher planet multiplicity and the differences in planet masses between systems, the similarity of the results suggests that similar dynamics are in play at the transition to instability.

To summarize, we determined an empirical stability limit for these \Kepler systems, finding that most of them have measured eccentricities that are lower than the limit of stability by a factor of a few to ten.
How might we interpret these results?
To understand if this distribution of systems could be explained by dynamical sculpting through giant impacts, we need to compare it with theoretical predictions from planet formation theory.

\section{Comparison with Giant Impact Formation Hypothesis \label{sec:giant_impact}}

\begin{figure}
\epsscale{1.1}  
\plotone{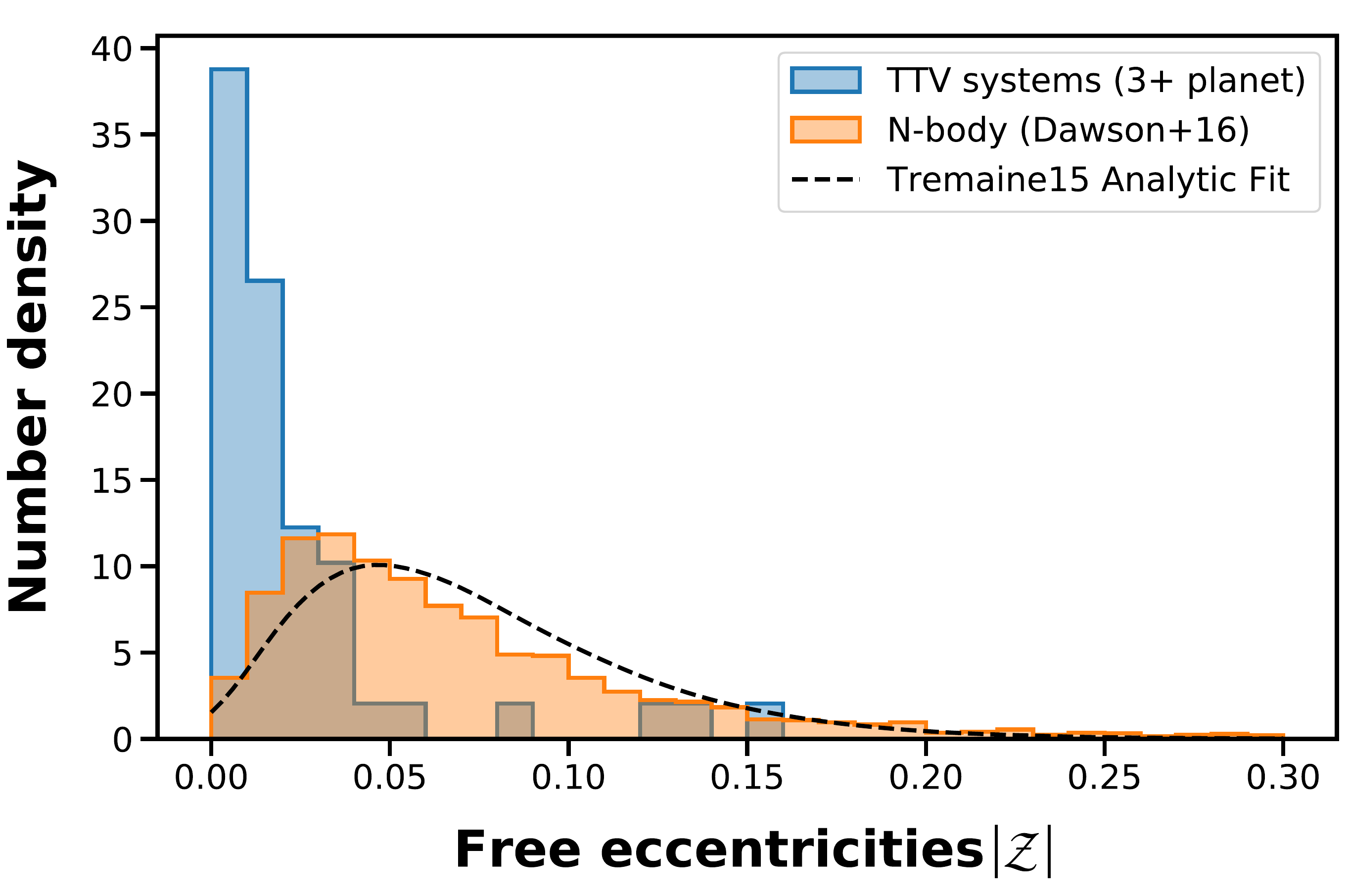}
\caption{\added{The magnitude of the} measured free eccentricities $\abs{\ZZ}$ of the TTV systems in our sample (blue), compared with the outcomes of the $N$-body simulations from \cite{Dawson2016} (\texttt{Eh} suite of simulations) in orange.
The eccentricities resulting from $N$-body simulations are significantly higher than those in actual systems.
We find that the analytic distribution from \cite{Tremaine2015}, shown as a dashed line, is able to fit the $N$-body results reasonably well.
\label{fig:dawson_z_comparison}
}
\end{figure}

In the giant-impact formation hypothesis, the final stage of terrestrial planet growth occurs through the mutual collision of the largest available bodies.
One prediction of this theory is that in the absence of any further damping processes, planetary systems should be left in a dynamically excited state.
Numerous authors (e.g. \citealt{Chambers1998,Chambers2001,Raymond2005,Obrien2006,Hansen2013,Dawson2016,Izidoro2017,Mulders2020}) have performed $N$-body simulations of this phase of planet formation.
% \memodt{We should include a couple references to the older literature on this. I think there are several Chambers papers on this in the early 2000s, but I don't think 1996 is the right reference}.
These simulations have often been able to reproduce some properties of observed planetary systems by varying the initial conditions, such as disk surface density, or the spacing, eccentricity and inclination distributions of the starting embryos.

As a representative example, we looked at the results from the \texttt{Eh} suite of simulations from \cite{Dawson2016}.
In this numerical experiment, planet embryos were spread between $\approx 0.05-1$ AU, given small initial eccentricities and inclinations, and allowed to evolve for 27 Myr in the absence of any dissipation from gas.
The systems undergo dramatic rearrangements at erratic intervals through collision and scattering events, so the resulting systems should have a wide range of eccentricities, up to the stability limit.
We computed approximate free eccentricities, $\ZZ \approx (z' - z)/\sqrt{2}$, for each pair of adjacent planets in the final simulation results.
These are shown in Figure \ref{fig:dawson_z_comparison}.
We find that the simulated planets have significantly larger eccentricities than the measured eccentricities of the \Kepler TTV planets in our sample.

This comparison is complicated by the fact that the outcomes of the $N$-body simulations only resemble the observed systems in some ways, and differ in others.
For example, the \texttt{Eh} simulations from \cite{Dawson2016} are able to reproduce the period and period ratio distributions of the observed \Kepler multi-planet systems in broad terms, but the simulated systems typically have much larger numbers of transiting planets than are observed.
Furthermore, to make the comparison with our sample of \Kepler TTV systems, we would also need to focus only on $N$-body outcomes where the resulting planets are closely-packed and close to MMRs.
This is important since MMRs strongly mold stability boundaries \citep{Obertas2017}, so comparing our sample to a broader population that includes systems far from resonance may be misleading.
Ultimately, any comparison between observed systems and the results of $N$-body simulations will be complicated by particular choices in the initial conditions and any filtering of the outcomes.
Considering that the parameter space for such simulations is large and has not been fully explored, we turn instead to a more general theory to make a different comparison to the observed systems.

\subsection{The Ergodic Hypothesis}

\cite{Tremaine2015} used the tools of statistical mechanics to predict the distribution of orbital elements of planets following the giant-impact phase.
He posited that in the absence of further dissipation, mutual gravitational interactions between planets allow systems to explore all of the stable phase space available to them, which he termed the ergodic hypothesis.
From this ansatz, \cite{Tremaine2015} found analytic distributions for the eccentricities and semimajor axes in coplanar systems as a function of a single free parameter: the dynamical temperature $\tau$.
He showed that these results were able to match the outcomes of $N$-body giant impact simulations by \cite{Hansen2013}.
We find similarly good agreement between the ergodic hypothesis and the gas-free simulations of \cite{Dawson2016}.
Figure \ref{fig:dawson_z_comparison} shows the analytic distribution from \cite{Tremaine2015}, fitted to the simulation outcomes by optimizing the dynamical temperature. 
Here we adopt the approach of \cite{Tremaine2015} with two main modifications.

First, \cite{Tremaine2015} uses a simplified stability criterion to make analytic progress, requiring that the distance between each planet's apocenter and its outer neighbor's pericenter exceed a fixed number of mutual Hill radii.
While the good agreement between the model's predictions and $N$-body simulations with uniformly distributed planetary embryos suggests this criterion is useful on average, one would expect it to change significantly near MMRs, where instability times change by several orders of magnitude \citep{Obertas2017}.
Given that all our TTV planets are near MMRs, we adopt the ergodic ansatz of \cite{Tremaine2015}, but rather than taking his analytic integrals over phase space using a simple stability criterion, we evaluate them numerically using SPOCK to assess stability.

Second, in addition to modeling orbital eccentricities, \cite{Tremaine2015} considers all possible interplanetary separations to arrive at a distribution of expected planetary spacings. 
In our case, we consider specific systems where the orbital periods (and thus the spacings) are precisely measured by the transit data.
We therefore choose to evaluate the expected distribution of orbital eccentricities conditioned on the precisely measured orbital periods. 

\subsection{Experimental Setup} \label{sec:ergsetup}
\begin{figure*}
\plotone{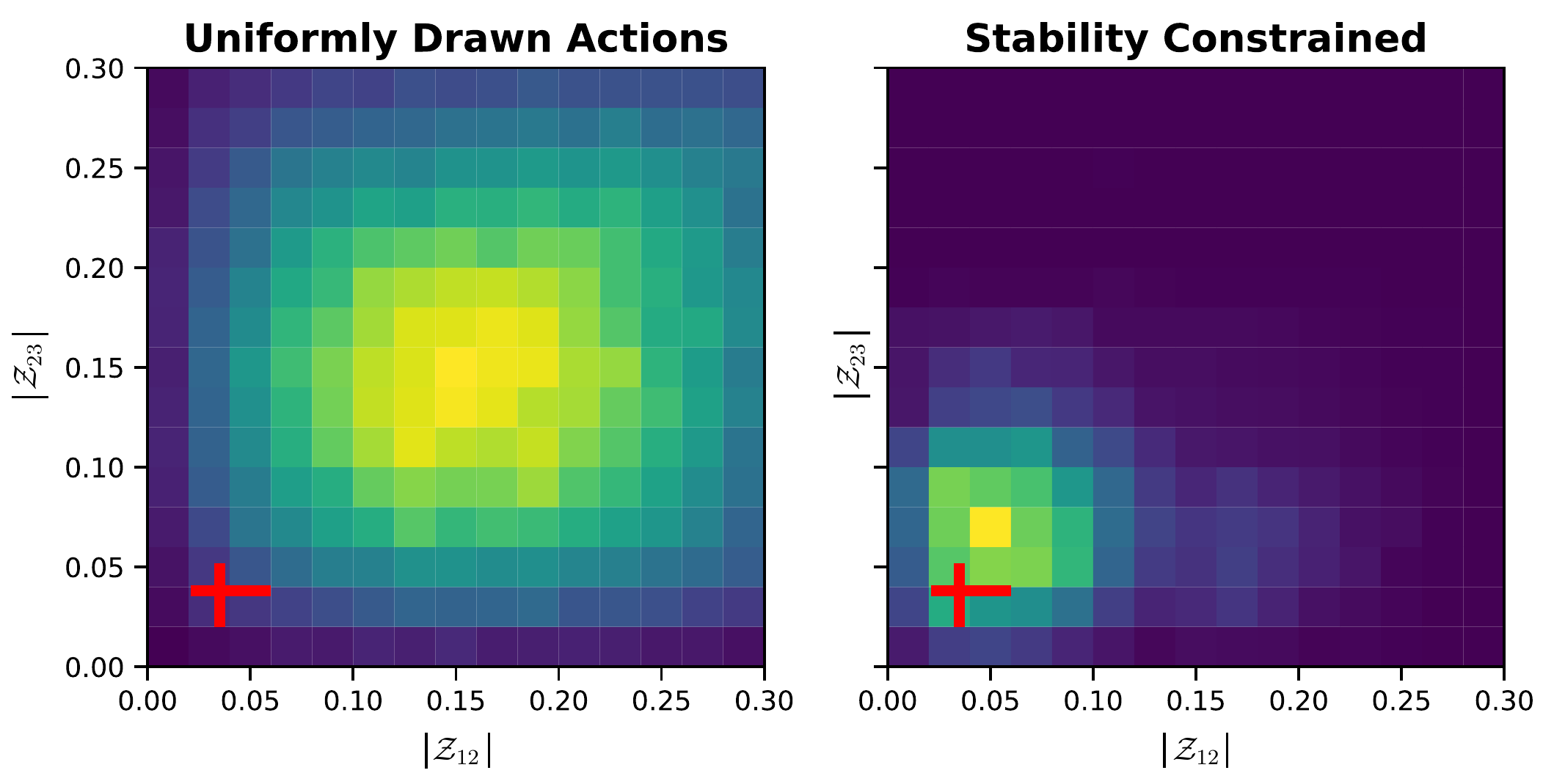}
\caption{
An example stability map for Kepler-24, a three-planet system.
Here, we show the \added{magnitude of the} combined eccentricities $\abs{\ZZ}$ for the two adjacent planet pairs, where the red cross denotes the measured values from H17.
\newline
\textit{Left}: Drawing actions uniformly without concern for stability, results in a distribution of $\abs{\ZZ}$ biased toward large values, with the position of the peak determined by the upper eccentricity cutoff $e_\mathrm{cross}$.
\textit{Right}: When the eccentricity distribution is weighted by the probability of stability returned by SPOCK, it shifts to much lower eccentricities, with the high-eccentricity configurations excluded by stability.
\label{fig:stability_map}
}
\end{figure*}

The ergodic hypothesis of \cite{Tremaine2015} posits that for a fixed set of input masses for planets on coplanar orbits, the orbital actions will be uniformly distributed over the region of phase space that is dynamically stable. 
As mentioned above, given TTV systems' near resonant configurations, we fix the planets' actions (per unit mass) $\Lambda = \left(GM_\star a\right)^{1/2}$, where $M_\star$ is the stellar mass, $a$ is the semimajor axis, and $G$ is the gravitational constant.
These values are precisely determined by the observed orbital periods to within an overall constant that does not affect the calculation.
This leaves one remaining action for each planet, $\Gamma = \sqrt{GM_\star a}(1-\sqrt{1-e^2}) \approx \sqrt{GM_\star a}e^2/2$.
We draw $\Gamma$ for each planet independently and uniformly, and the mean anomalies and longitudes of pericenter (conjugate to the $\Lambda$ and $\Gamma$ actions, respectively) uniformly from $[0, 2\pi]$.
In order to limit the number of stability evaluations, we impose a minimal criterion for stability by limiting the $\Gamma$ actions to a maximum eccentricity of $e_\mathrm{cross} = (a' - a)/a$ (c.f. Eq. \ref{eq:Zcross}).
As in Section \ref{sec:alpha_ecc}, in order to marginalize over the planetary mass uncertainties, for each draw we sample the planetary masses from the posterior distribution of H17.

As an example, in the left panel of Fig.\:\ref{fig:stability_map} we show this distribution for the 3-planet system Kepler-24, which we found in the previous section was the system closest to its stability limit, with $\abs{\ZZ/\Zunstable} \approx 0.8$.
In order to visualize this 3-D distribution, we have projected it onto the 2-D space spanned by each adjacent pair of planets' free eccentricities \ZZ, which are the combinations precisely measured by TTV observations.
Since eccentricities are drawn uniformly in $\Gamma \propto e^2$, configurations are skewed toward high values of $\abs{\ZZ}$, with a peak determined by the cutoff value of $e_\mathrm{cross}$ in our sampling procedure\footnote{The largest $\abs{\ZZ}$ values, obtained by drawing both eccentricities near $e_\mathrm{cross}$ in an {\it anti-aligned} configuration, are disfavored by the requirement that the randomly drawn pericenter longitudes differ by $\pi$.}.

Finally, we use SPOCK to evaluate a probability of stability for each input configuration over $10^9$ orbits.
Rather than labeling each configuration as definitely stable or unstable, we weight each configuration by the probability of stability estimated by SPOCK, which provides a smoother and more reliable result \citep{Tamayo2021}.
The result is shown in the right panel of Fig.\:\ref{fig:stability_map}, 
where we see that stability restricts the available phase space down to significantly lower free eccentricities.
This represents the giant-impact phase prediction as evaluated through the ergodic ansatz.
In other words, it is the approximate distribution of free eccentricities we would expect had we run many giant impact $N$-body simulations, and only retained the outcomes consistent with the observed orbital periods and planetary masses.

We compare this giant-impact phase prediction to the free eccentricities measured observationally by H17, marked by the red cross in Figure \ref{fig:stability_map}.
We find in this case that the peak of the predicted eccentricity distribution occurs at roughly $1.5\times$ the measured values for the Kepler-24 system.

\subsection{Results} \label{ssec:ergodic_results}

\begin{figure}
\epsscale{1.2}
\plotone{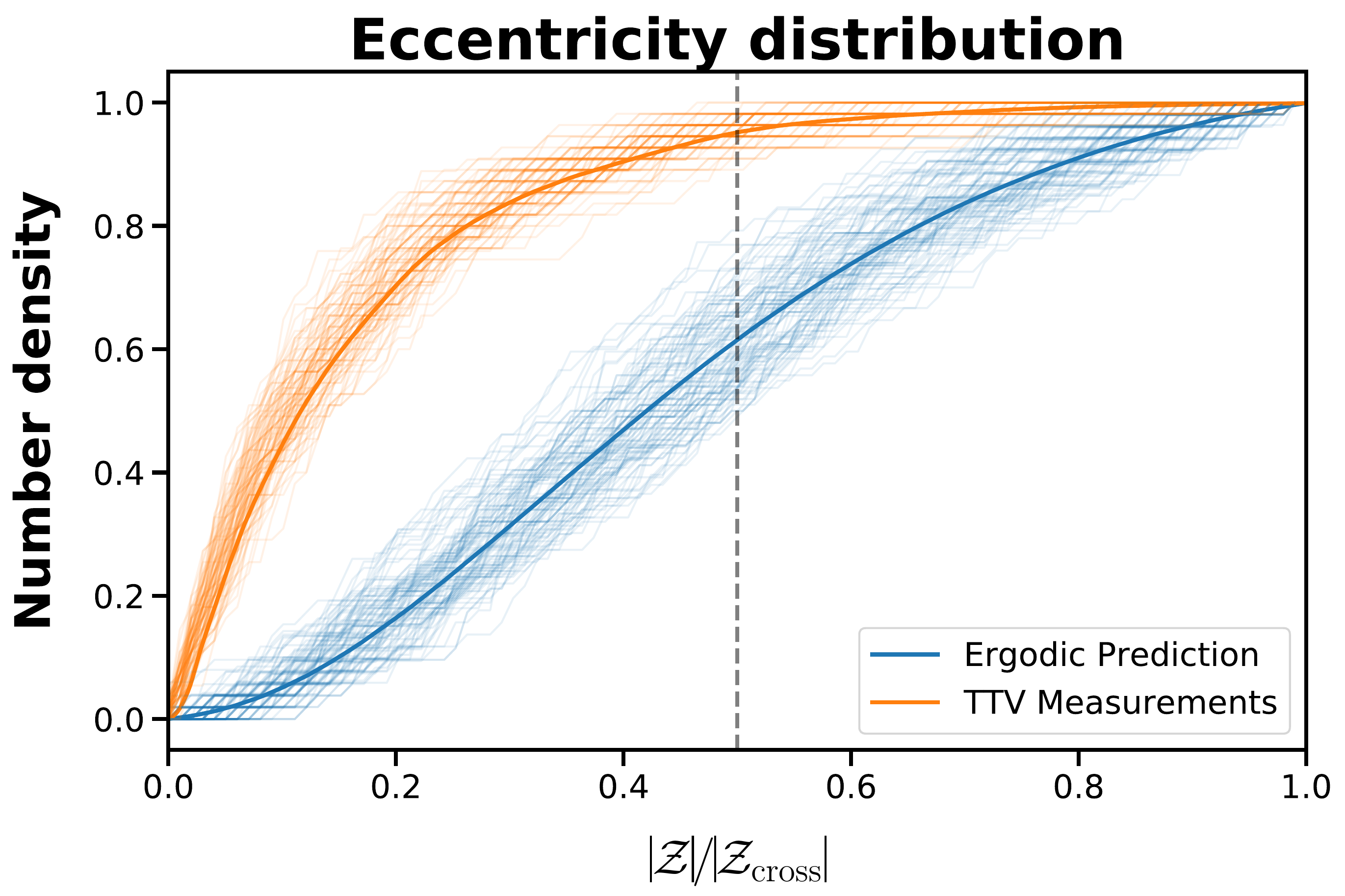}
\caption{Cumulative histogram of $\abs{\ZZ}/\abs{\Zcross}$ when drawn from the ergodic probability distributions (blue), compared with the same quantity measured for the TTV systems (orange).
The solid lines depict the cumulative distribution over all systems and draws, while the faded lines show each individual draw of the planet population, giving a sense of the width of the respective distributions. 
\label{fig:Z_Zcross_cumulative}
}
\end{figure}

We repeat the procedure in Sec.\:\ref{sec:ergsetup} for each of our 19 TTV systems.
Because we now have a multi-dimensional and probabilistic stability limit, we cannot measure a single fractional distance to instability, as in Section \ref{sec:alpha_ecc}.
Instead, we analyze the population as a whole, to compare the observed eccentricities with the predictions.

For each of the 19 \Kepler TTV systems in our sample, we draw free eccentricities \ZZ for each planet pair from the predictions of the ergodic hypothesis (e.g. the right panel of Fig. \ref{fig:stability_map}), where we have smoothed the probability distributions from our sampling procedure with a Gaussian kernel density estimate (KDE).
To put the eccentricities from each planet pair on the same scale, we normalize them by their respective $\abs{\Zcross}$.
This yields a synthetic population of 49 planet pairs that have the same semimajor axes as the observed systems in our sample, but eccentricities generated via the ergodic hypothesis.
We then repeat this procedure, generating 1000 sets of synthetic planet populations.

This can be compared with the eccentricities measured by TTVs.
To account for measurement uncertainties, we generate a synthetic `observed' population by drawing a single sample from the posterior distributions of each system from H17.
We then repeat this process to generate 1000 observed populations.

We show the results in Figure \ref{fig:Z_Zcross_cumulative}.
The solid central lines denote the cumulative $\abs{\ZZ}/\abs{\Zcross}$ distribution over all systems and draws, while each fainter line shows the cumulative distributions for a single synthetic population.
It is visually obvious that the two distributions are different, as confirmed with a Kolmogorov-Smirnov test ($p < 10^{-9}$).
The observed eccentricities are far below what would be expected under the ergodic hypothesis, i.e., smaller than would be expected from a giant-impact phase of planet formation.
Under the hypothesis that planetary systems are evenly distributed throughout the stable phase space available to them, $\sim 40\%$ of planets should have eccentricities more than halfway to orbit crossing (as can be seen from the vertical dashed line in Figure \ref{fig:Z_Zcross_cumulative}).
In the actual observed population, only a few planets have such high eccentricities.
Most systems have eccentricities $\sim 5\times$ below orbit-crossing values.
The median observed eccentricity for the TTV planets is $\abs{\ZZ}/\abs{\Zcross} \approx 0.1$, compared with a median of $\abs{\ZZ}/\abs{\Zcross} \approx 0.45$ from the ergodic hypothesis.

\added{
\subsection{Dissipation} \label{ssec:dissipation}
\begin{figure}
\epsscale{1.2}
\plotone{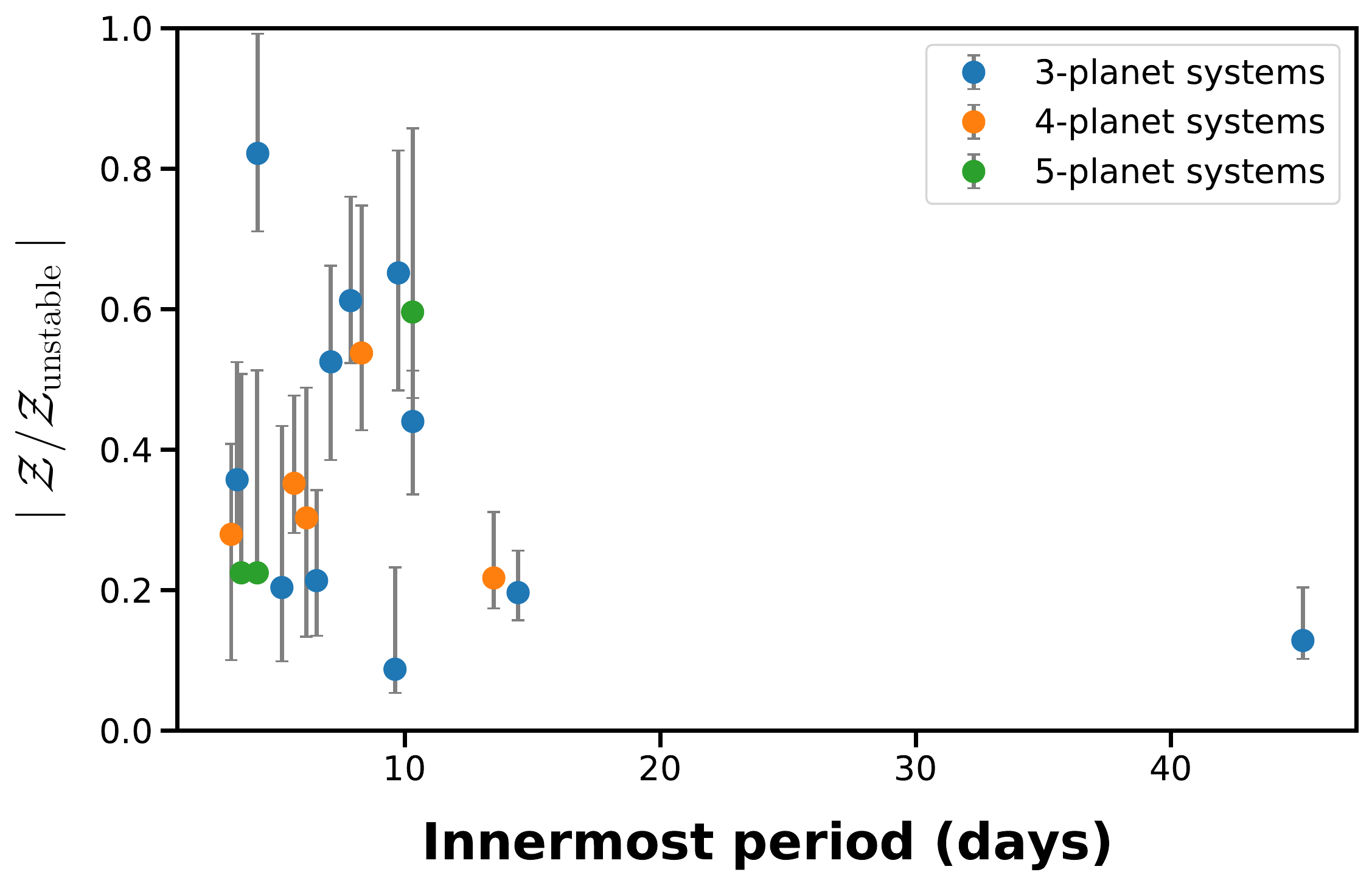}
\caption{The fractional distance to instability for each system, $\abs{\ZZ/\Zunstable}$, as a function of the shortest orbital period in the system.
Here, we have used the period of the innermost known planet, even if that planet did not show TTVs.
There is no clear trend, whereas tidal dissipation predicts that the systems with the closest-orbiting planets also be the most strongly damped.
\label{fig:alpha_per_innermost}
}
\end{figure}

This finding that the observed planets have eccentricities significantly lower than the expectations from the ergodic hypothesis leads us to conclude that if these systems did form through a giant-impact phase, the eccentricities must have been damped between then and now.

In studying terrestrial planet formation within the Solar System, \cite{Obrien2006} found that a more realistic simulation of dynamical friction between planet embryos and planetesimals could damp the eccentricities and inclinations of the planets formed to be a better match with the observed orbits of the Solar System planets.
\cite{Mulders2020} obtained similar results in their simulations of exoplanet systems, finding that including 25\% of the system mass as planetesimals ($M \sim 0.005 M_\oplus$) in their $N$-body simulations could better reproduce the eccentricity and inclination distributions of the \Kepler population, as compared with simulations where all the system mass was contained in the planet embryos.

Leftover gas from the protoplanetary disk phase may also be responsible for this damping.
For example, \cite{Dawson2016} found that including residual gas corresponding to $\approx$1\% of the gas density in the minimum-mass solar nebula for the first 1 Myr of their $N$-body giant-impact simulations produced systems with significantly lower final eccentricities that were more similar to observations.
This stands in contrast to their gas-free simulations (\texttt{Eh} suite discussed in Figure \ref{fig:dawson_z_comparison}), in which the resulting systems had much larger eccentricities.

Alternatively, the giant-impact process itself may be self-regulating.
\cite{Kobayashi2019} argued that damping from the planetesimal disk may be inefficient because the planetesimals are quickly ground down to smaller sizes in a collisional cascade.
However, a giant impact between two planet embryos also generates a swarm of ejecta on similar orbits to the combined protoplanet, which then scatters this ejecta into the outer disk.
These authors found that this process could reduce the planet's eccentricity by an order of magnitude.
\cite{Poon2020} also investigated the effect of imperfect accretion in the giant-impact formation scenario, but their simulations produced planets whose eccentricities were too damped compared with observations.

It is also possible that eccentricities are damped over longer timescales.
\cite{Hansen2015} found that as the innermost planets in a multi-planet system are subject to orbit circularization and decay from tidal dissipation due to time-variable tidal bulges raised by the star, secular interactions with the outer planets could lead to the circularization of all the planets in the system.
For a terrestrial-sized planet with an Earth-like tidal quality factor and an orbital period shorter than $\lesssim$20~days, the tidal dissipation timescale can be a few Gyr or less, comparable to the typical age of our systems.
Most of the systems in our sample have at least one planet in this orbital period and radius range, and may therefore undergo this process, subject to the usual large uncertainties in the tidal quality factor and its dependence on forcing frequency.

To investigate this, we plot the fractional distance to instability for each system, against the period of the innermost planet (Figure \ref{fig:alpha_per_innermost}).
From our relatively small sample, there is no obvious trend.
If tides indeed dominate the dissipative histories of \Kepler multi-planet systems, then the systems hosting the closest-in planets should be farthest from the stability limit due to the stronger tidal dissipation \citep[e.g.,][]{Delisle2014}, while those without such close-in planets should retain their large, primordial eccentricities from the giant impact phase.
This is not seen in Fig. \ref{fig:alpha_per_innermost} -- in particular, the rightmost systems in that plot are also the furthest from the stability limit.
Unless those systems host undetected planets at shorter orbital periods, it is unlikely that tidal dissipation played a major role in damping their eccentricities.
}

\section{Discussion \label{sec:discussion}}

In the previous two sections, we performed numerical experiments to show that our sample of \Kepler TTV systems has eccentricities significantly lower than would be expected from a dissipation-free giant impact phase.
\added{While our results cannot differentiate between the various dissipative processes outlined in Section \ref{ssec:dissipation}, they show that at least some form of damping is required if these planets indeed formed through a late stage of giant impacts.}
While the use of the machine-learning tool SPOCK to speed up stability classification allowed us to explore the phase space in greater detail than in previous work relying on $N$-body integrations, we now consider the validity of several of our assumptions.

\subsection{Timescale of Stability}

\begin{figure}
\epsscale{1.2}
\plotone{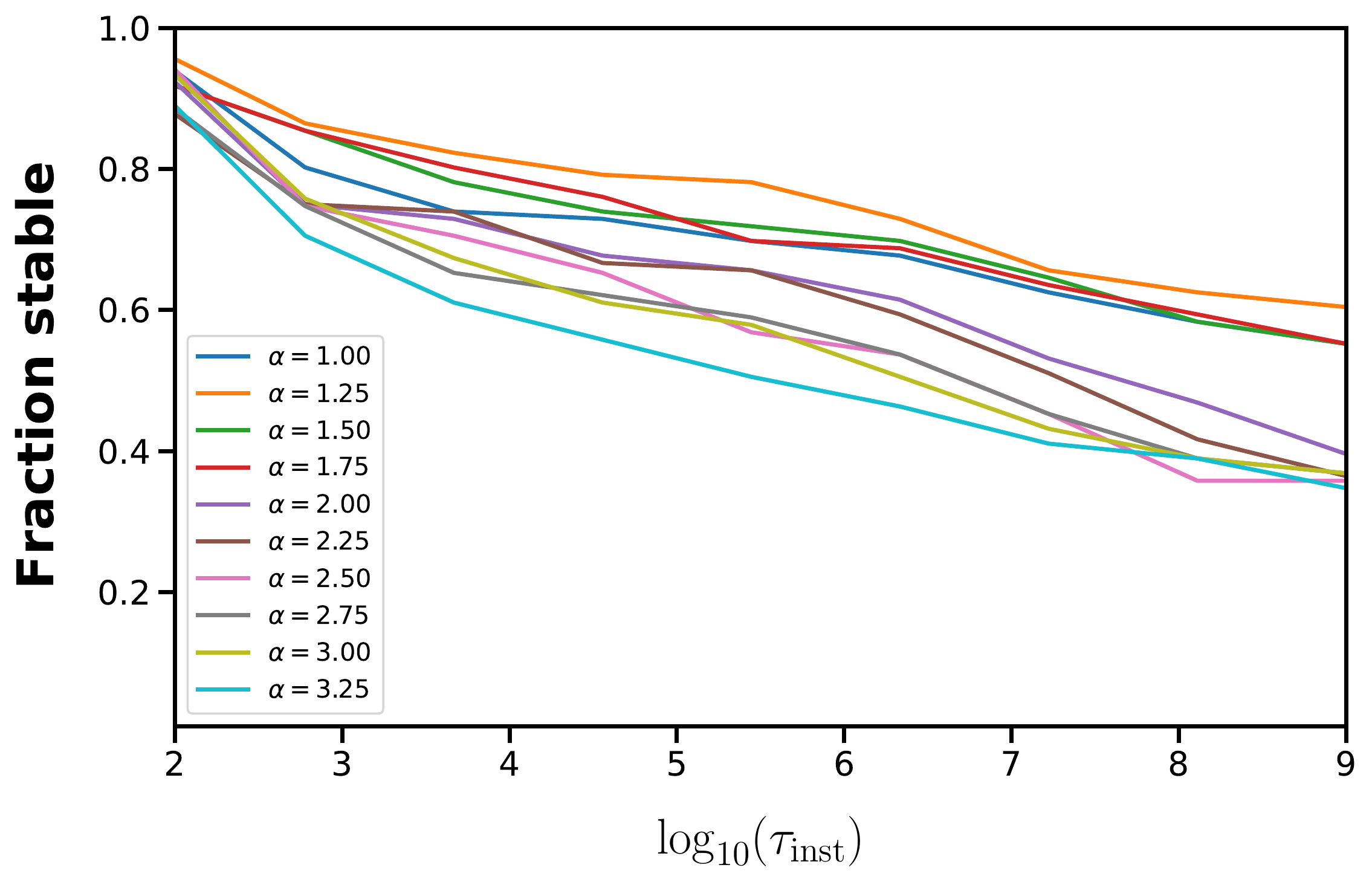}
\caption{Fraction of initial configurations stable out to a given instability time, determined by $N$-body integrations.
This plot shows the results for the Kepler-52 system, with 100 sets of initial configurations each with their eccentricities inflated by the listed values of $\alpha$.
Similar to \cite{Volk2015}, we find that the fraction of systems that go unstable is roughly constant per decade in time, allowing us to approximately extrapolate to $10^{11}$ orbits.
\label{fig:instability_time}}
\end{figure}

SPOCK was trained to classify the stability of planetary systems over $10^9$ orbits of the innermost planet.
For the systems we have examined, the innermost planet has a period of a few to tens of days, so when SPOCK classifies a system as ``stable'', stability is only guaranteed for a few tens of millions of years, which is much shorter than typical system ages of several billion years.
Requiring that systems be stable over longer timescales than this would move our numerically-determined stability limits to lower eccentricities.

However, we do not expect that extending the stability requirement to even $10^{11}$ orbits would shift our predicted stability envelopes enough to affect our conclusions.
While repeating the entire experiment with $N$-body integrations would be computationally infeasible, we performed some limited comparisons.
We repeated the experiment of Section \ref{sec:alpha_ecc} for two of our systems, Kepler-52 and Kepler-24, where we increased all eccentricities in a system by a fixed factor $\alpha$, but using the \texttt{WHFast} $N$-body integrator \citep{Rein2015a} in the \texttt{REBOUND} package \citep{Rein2012}.
We integrated out to $10^9$ orbits of the innermost planet, recording whether each initial configuration survived to the end of the integration, and storing the time of instability if it did not (these results agreed well with those from SPOCK in Section \ref{ssec:alpha_results}).

In Figure \ref{fig:instability_time}, we show the fraction of initial configurations which remained stable out to a given time, for each value of the eccentricity inflation factor $\alpha$.
We find that the fraction of systems that goes unstable in a given decade of time is roughly constant, a result previously noted by \cite{Volk2015,Volk2020}.
Even for the most eccentric systems (highest $\alpha$ factors), only roughly $5-7\%$ of systems go unstable in each decade.
Extrapolating our results from $10^9$ to $10^{11}$ orbits, we would therefore expect a further $10 - 15\%$ of initial configurations to go unstable.

We should therefore expect that the eccentricity threshold for instability, \Zunstable, would be shifted by only a similar amount, $10-15\%$. 
Given that most of our systems are overstable by a factor of a few to ten, 
this suggests that even requiring systems to remain stable over their entire lifetime would not significantly affect our conclusions.

\added{
\subsection{Stability from Dissipation} \label{ssec:dissipation_stability}
SPOCK was trained on $N$-body integrations in the absence of any dissipative forces, so the stability classifications it makes are also predicated on a dissipation-free environment.
As discussed in Section \ref{ssec:dissipation}, tidal dissipation may be important for many of the systems in our sample, which have planets on close-in orbits.
\cite{Tamayo2015} showed numerically that tidal damping may stabilize systems when the tidal eccentricity damping timescale is shorter than the instability time.
Thus, some configurations classified by SPOCK as unstable over over $10^9$ orbits may actually be indefinitely stable due to tidal damping.

We believe that this should not have have a significant impact on our eccentricity limits for stability, however.
For example, if the tidal damping timescale were $10^8$ orbits, then SPOCK would misclassify systems with instability times between $10^8$ and $10^9$ orbits, as these diffuse slowly enough for tidal damping to stabilize the system.
However, stability has a steep dependence on eccentricity, as we showed in Figure \ref{fig:linear_alpha_example} (in terms of the SPOCK probability rather than the direct instability timescale).
Thus one only has to slightly increase the eccentricities beyond our estimated values before the instability times drop below $10^8$ orbits, and tides are no longer able to stabilize the system.
In Figure \ref{fig:linear_alpha_example}, we argued that our particular choice of the SPOCK probability threshold was not important due to this steepness of the eccentricity dependence, and found that our computed stability limits did not change by more than 15\% even when we varied that threshold.
Here, we expect that the impact of tidal damping on our eccentricity limits would be similar, as tidal damping simply reduces the instability time threshold compared with SPOCK's $10^9$ orbits.
Hence, the additional stability conferred by tidal dissipation is unlikely to significantly affect our conclusions.
}

\subsection{Additional Planets} \label{ssec:additional_planets}
Another concern arises from the presence of additional planets not included in our analysis.
As mentioned in Section \ref{sec:systems}, six of the systems in our sample contain known additional planets or planet candidates which did not display TTVs.
In all of these cases, the additional planets were well-separated from those included in the analysis, with $P'/P > 2.2$, compared with $P'/P \leq 2.0$ for the TTV planets.
As the planets' eccentricities are increased, it is more likely that the more closely-packed TTV planets would drive the instability, compared with interactions with a more distant planet.

To check this assumption, we repeated our analysis in Section \ref{sec:alpha_ecc} for those six systems, while including the additional non-TTV planets.
Because these non-TTV planets or planet candidates only have radius measurements, and not mass measurements, we use the probabilistic mass-radius relationship from \cite{Ning2018} to assign them masses.\footnote{We utilize the Python package \texttt{MRExo} by \cite{Kanodia2019} to propagate uncertainties in the radius measurements and mass-radius relationship to uncertainties in planet mass.}
They were assigned eccentricities from a Rayleigh distribution with scale factor $\sigma_e = 0.06$, consistent with the population-level estimate for multi-planet systems from \cite{VanEylen2019}.
We find that the threshold for stability shifts by $\lesssim 5\%$, suggesting that the instability is primarily driven by the more closely-packed TTV planets.
For consistency among all systems and due to the large mass uncertainties, we did not include the non-TTV planets in our results above.

Apart from additional close-in, small planets, these \Kepler systems are also likely to have unseen, distant, giant planet companions (e.g. \citealt{Zhu2018b}).
While this would modify the secular dynamics, the stability of closely packed systems is largely set by the dominant interactions between nearby planets \citep{Tamayo2020}.
Barring extreme cases like large-scale instabilities between outer giant planets \citep[e.g.,][]{Huang2017}, we would not expect such distant giant planets to significantly change our results.
Unseen planets {\it between} observed TTV planets would strongly modify our conclusions, but such cases should be rare.
\cite{Gilbert2020} used an information-theoretic framework to infer that $\sim 20\%$ of multi-transiting systems could contain undetected planets intermediate to the known planets.
However, such systems would be observed to have a larger spacing between transiting planets of $\sim 30$ Hill radii, whereas the TTV systems we have examined are all more closely packed than that.

\subsection{Inclinations} \label{sec:inc}
We performed our stability classifications assuming that all systems are coplanar.
Indeed, theoretical and numerical work suggests that small inclinations do not play a major role in the stability of compact systems \citep{Wisdom1980a,Quillen2011,Obertas2017,Hadden2018,Petit2020}. 
\cite{Tamayo2020} experimented with inclination-related features in SPOCK but found no performance improvement, empirically suggesting that inclinations can be ignored at first approximation for roughly coplanar systems ($i \lesssim 10^\circ$), as is the case for our multi-transiting systems.

Mutual inclinations between planets may also affect the detectability of the system.
By examining only planetary systems containing multiple transiting planets, we are biasing our analysis to systems with low mutual inclinations.
If eccentricities and inclinations are excited similarly, then this means our sample of multi-transiting systems is also biased toward low eccentricities.
We check on the possible selection effects on our results in Section \ref{ssec:ergodic_results}, by assigning each planetary configuration an inclination equal to its eccentricity, in radians, as motivated by studies of viscous stirring of planetesimals \citep{Kokubo2005,Dawson2016}.

We then use the CORBITS code \citep{Brakensiek2016} to evaluate the probability that all planets transit in each of our orbital configurations.
The eccentricity distributions generated by the ergodic hypothesis were then re-weighted according to this multi-transit probability.
We found that adding this constraint did not significantly shift the peak of those eccentricity distributions, i.e. the requirement of dynamical stability was always more constraining than the geometric requirement for all planets to transit.

\subsection{Applicability to Non-TTV Systems}
We restricted our analysis to only \Kepler TTV systems, which have well-measured planet masses and eccentricities.
This may be a dynamically special subset of systems, which are close to mean-motion resonances.
We found that in this sample, the observed eccentricities do not bring the systems close to the edge of stability, but instead have significant room to grow.
\cite{VanEylen2019} used transit durations to find that the entire population of \Kepler multi-transiting systems appear to have larger eccentricities $\sigma_e \approx 0.06$ compared with a sample containing just the TTV systems $\sigma_e \approx 0.02$.
If this were true at the individual level, then such systems may be closer to the stability limit than the ones we have analyzed.
On the other hand, the non-TTV systems have typically larger spacings between planets and are farther from MMRs, so it may be possible to increase their eccentricities to larger values before reaching the edge of instability.
Future work could use SPOCK to map out stability boundaries for all \Kepler systems.
However, such a study would differ from ours because the actual eccentricities of most Kepler systems are poorly constrained.

\section{Conclusion \label{sec:conclusion}}
In this study, we investigated whether compact multi-planet systems have eccentricities close to the stability limit.
We utilized SPOCK, a machine-learning model from \cite{Tamayo2020}, to predict the long-term stability of planetary configurations as we varied their eccentricities.
We focused on \Kepler TTV systems with well-measured planet masses and eccentricities, which would facilitate comparison with the threshold for stability.

We increased all the free eccentricities within each system by a constant factor, monitoring the effects on long-term dynamical stability.
The results showed that while a few systems are indeed close to instability, most could withstand having their eccentricities increased by a factor of a few before going unstable.
To put these results into context, we compared them with predictions from planet formation theory.
If these compact systems form through a phase of giant impacts after the gas disk has dissipated, the eccentricities of planets should be distributed throughout the available phase space up to the limits of stability, as suggested by \cite{Tremaine2015}.
We found that such a distribution is inconsistent with observations, as stability permits eccentricities substantially larger than those observed.
A population-level analysis found that the TTV systems in our sample have a median eccentricity about 4 times lower than the median eccentricity we might expect if those same planets were generated through a process described by the ergodic hypothesis.

These findings suggest that if terrestrial planets indeed form through giant impacts, then an additional source of damping is required to explain their low, yet non-zero, observed eccentricities.
\replaced{
This is consistent with the results of some $N$-body simulations of the giant impact phase of planet formation like those by \cite{Dawson2016,MacDonald2020}, who found that some residual gas at the beginning of this phase better reproduces certain properties of the observed \Kepler planets.
Further theoretical and observational work could help pin down the initial conditions for planet formation and provide a mechanism by which such damping may operate, for instance through interaction with a remnant gas disk or a planetesimal disk.}
{This may be the result of a dissipative formation environment, either due to residual gas at the beginning of the giant-impact phase (e.g. \citealt{Dawson2016,MacDonald2020}, or through dynamical friction between planetary embryos and planetesimals \citep{Obrien2006} or collisional ejecta \citep{Kobayashi2019,Poon2020}.
Alternatively, damping of the planets' eccentricities may occur over the age of the system, driven by tidal dissipation in the innermost planet(s) and secular coupling with the rest of the system \citep{Hansen2015}.
In the latter case, the eccentricity distributions in multi-planet systems should evolve on the $\sim$Gyr timescales of tidal eccentricity damping, a hypothesis potentially testable using the many new planetary systems being discovered by \textit{TESS}.
Our results do not differentiate between these possiblities, which we leave for future work to explore.
}

This work also demonstrates the utility of machine-learning tools like SPOCK, which enabled the classification of large numbers of planetary configurations that would otherwise be prohibitively computationally expensive with traditional $N$-body integrations.
Such models open up new ways to explore the complex phase space and phenomena of planetary dynamics.

\acknowledgements

We thank Rebekah Dawson for providing the results of the simulations from \cite{Dawson2016}.
\added{
We also thank the referee for helpful comments that improved the manuscript.}

\software{
\texttt{SPOCK} \citep{Tamayo2020};
\texttt{rebound} \citep{Rein2012,Rein2015a};
\texttt{CORBITS} \citep{Brakensiek2016};
\texttt{MRExo} \citep{Kanodia2019};
\texttt{celmech} (\url{https://github.com/shadden/celmech});
\texttt{astropy} \citep{Astropy13,Astropy18};
\texttt{numpy} \citep{Numpy};
\texttt{scipy} \citep{Scipy};
\texttt{matplotlib} \citep{Matplotlib}.}

\bibliography{manuscript,software}
\bibliographystyle{aasjournal}

\end{document}

%% file: tables/table1_stub.tex
% planet, period, Mstar, Mpl, Zd, Zcom, Z_Zunstable
Kepler-11 b & 10.304 & $0.9^{+0.1}_{-0.1}$ & $0.7^{+0.3}_{-0.2}$ & $0.028^{+0.010}_{-0.006}$ & $0.007^{+0.004}_{-0.005}$ & $0.60^{+0.26}_{-0.12}$ \\
Kepler-11 c & 13.025 & $0.9^{+0.1}_{-0.1}$ & $1.8^{+0.9}_{-0.5}$ & $0.013^{+0.009}_{-0.015}$ & $\cdots$ & $\cdots$ \\
Kepler-11 d & 22.687 & $0.9^{+0.1}_{-0.1}$ & $6.8^{+0.7}_{-0.8}$ & $0.009^{+0.001}_{-0.001}$ & $\cdots$ & $\cdots$ \\
Kepler-11 e & 31.995 & $0.9^{+0.1}_{-0.1}$ & $6.7^{+1.2}_{-1.0}$ & $0.018^{+0.004}_{-0.005}$ & $\cdots$ & $\cdots$ \\
Kepler-11 f & 46.686 & $0.9^{+0.1}_{-0.1}$ & $1.7^{+0.5}_{-0.4}$ & $\cdots$ & $\cdots$ & $\cdots$ \\
\hline
Kepler-23 b & 7.107 & $1.0^{+0.1}_{-0.1}$ & $1.3^{+1.3}_{-0.5}$ & $0.017^{+0.006}_{-0.026}$ & $0.014^{+0.033}_{-0.011}$ & $0.52^{+0.14}_{-0.14}$ \\
Kepler-23 c & 10.742 & $1.0^{+0.1}_{-0.1}$ & $2.2^{+2.8}_{-0.9}$ & $0.021^{+0.014}_{-0.013}$ & $\cdots$ & $\cdots$ \\
Kepler-23 d & 15.274 & $1.0^{+0.1}_{-0.1}$ & $<2.4$ & $\cdots$ & $\cdots$ & $\cdots$ \\
\hline